\documentclass{pasj00}

\begin{document}
\SetRunningHead{Fujii et al.}{SiO Masers Survey of the Inner Bar of the Galactic Bulge}
\Received{2005/11/24}
\Accepted{2006/03/26}

\title{SiO Maser Survey of the Inner Bar of the Galactic Bulge}

\author{Takahiro \textsc{Fujii},$^{1,2}$ 
        Shuji \textsc{Deguchi},$^{3}$  Yoshifusa \textsc{Ita},$^{4,5}$}
\author{Hideyuki \textsc{Izumiura},$^{6}$ Osamu \textsc{Kameya},$^{1, 7}$ 
        Atsushi \textsc{Miyazaki},$^{3,8}$
}

\and
\author{ Yoshikazu \textsc{Nakada}$^{4,9}$}
\affil{$^{1}$ VERA Project Office, National Astronomical Observatory, 
              2-21-1 Osawa, Mitaka, Tokyo 181-8588}
\affil{$^{2}$ Faculty of Science, Kagoshima University,
              1-21-35 Korimoto, Kagoshima, Kagoshima 890-0065}
\affil{$^{3}$ Nobeyama Radio Observatory, National Astronomical Observatory,\\
              Minamimaki, Minamisaku, Nagano 384-1305}
\affil{$^{4}$ Institute of Astronomy, School of Science, The University of Tokyo,\\
              2-21-1 Osawa, Mitaka, Tokyo 181-0015}
\affil{$^{5}$Institute of Space and Astronautical Science, Japan Aerospace Exploration Agency, 
\\ Yoshinodai 3-1-1, Sagamihara, Kanagawa 229-8510}
\affil{$^{6}$ Okayama Astrophysical Observatory, National Astronomical 
Observatory, \\ Kamogata, Asakuchi, Okayama 719-0232}
\affil{$^{7}$ Mizusawa Astrogeodynamics Observatory, National 
Astronomical Observatory, \\ 2-12 Hoshigaoka, Mizusawa, Iwate 023-0861}
\affil{$^{8}$ Shanghai Astronomical Observatory, Chinese Academy of Sciences \\
80 Nandan Road, Shanghai, 200030, P.R. China}
\affil{$^{9}$Kiso Observatory, School of Science, The University of Tokyo, 
\\ Mitake, Kiso, Nagano 397-0101}

\author{(PASJ 58, No. 3 in press ; Ver 1.3 -- May 1, 2006)}

\KeyWords{Galaxy:  center,  kinematics and dynamics --- masers ---
stars: AGB and post-AGB} 

\maketitle

\begin{abstract}

We surveyed 291 MSX/2MASS infrared objects in the
$7^{\circ} \times 2^{\circ}$ area of
the galactic center in the 43 GHz SiO $J=1$--0 $v=$ 1 and 2 maser lines,
obtaining accurate radial velocities of 163 detected objects. The surveyed area 
is the region where the IRAS catalog is incomplete due to contamination 
by high source density. The objects in the present MSX/2MASS sample 
were chosen to have similar infrared characteristics  
to those of the previous SiO-maser-survey samples based on the color selected IRAS sources. 
The sampling based on the
2MASS catalog causes a bias to the foreside objects of the bulge due to
heavy obscuration by interstellar dust;
the detections are considerably leaned on the $V_{lsr}<0$ side.
The $l$--$v$ diagram reveals two conspicuous features,
which were not present or tenuous in the previous studies: one feature
indicating a linear velocity increase with longitude with $|l|<1.5^{\circ}$, which is likely
associated with the inner bar, and the other feature having
considerably eccentric velocities more than those of the normal $x_1$-orbit family feature.         
The extinction-corrected $K$ magnitudes  (if used as a distance modulus) 
tend to show  a sequential deposition of these objects along the line of sight 
toward the Galactic center depending on their radial velocities.
The tendency that appeared in the distance measures is consistent with the bulge-bar 
dynamical model utilizing the periodic orbit families in the bar potential.

\end{abstract}

\section{Introduction}

The Galactic bulge embraces a large number of mass-losing 
Asymptotic Giant Branch (AGB) stars (\cite{hab87}).
They supply gases to the Galactic nuclear disk, gradually fueled 
for the activity of the Galactic center (\cite{bli93}). 
The bar structure of the bulge is requisite for infalling the gases
to the central nuclear cluster containing the central black hole (\cite{reg04}).
Though a considerable amount of theoretical and observational works have been made 
for understanding the central part of the Galaxy, 
it is still not very clear how the stars 
revolve around the Galactic center (GC) in the bar-like potential, 
and how they play in feedback loops of reincarnation 
through the star-forming gas clouds to the bar structure. 

In theories, stellar orbits in the bar-like bulges are mainly composed of the $x_1$ 
and $x_2$ families, which fabricate the bar with two elongated features 
along and perpendicular to the bar (\cite{con89a}; \cite{sel93}; \cite{bin98}). In the past studies,
the large-scale bar structure with $\sim$3:1 elongation, which 
appears to make an angle of $\sim 30^{\circ}$ from the Sun-GC direction,
revealed in the distribution of infrared sources (\cite{nak91}; \cite{bli91}; \cite{dwe95,nik97};
\cite{bab05}),
and other objects (\cite{pat94}). Moreover, 
the position-velocity diagrams of the CO, $^{13}$CO, and HI gas distributions
clearly indicate the presence of such inner-bar structure (\cite{bin91,wei99}).
However, the inner structure corresponding to
the $x_2$ orbits did not clearly reveal in the near-infrared COBE maps (
\cite{bin96}; \cite{lau02}),
Recently, \citet{ala01} and \citet{nis05} analyzed the NIR photometric
data of stars toward the inner galactic bulge, finding a presence 
of the small-scale feature ($|l|<2^{\circ}$) which is possibly elongated
perpendicularly to the large-scale bar structure.
 
In order to study kinematics of stars in the Galactic bulge 
and its relation to the bar dynamics, we performed in this paper 
a new survey of mass-losing infrared objects in the central part of the Galaxy 
using SiO maser lines. The SiO maser lines give a precise radial velocity of
the central star under mass loss, i.e., the mean velocity of approaching and 
receding parts of the circumstellar shell (\cite{jew91}). Therefore,
obtained radial velocities can directly be interpreted as the 
stellar velocities. Because ages of these AGB stars are $\sim 0.2$--1 Gyr (\cite{mou02}),
they are dynamically relaxed in the bulge (\cite{ara05}), and therefore
can be used as good tracers of the velocity field of the bar-like bulge. 

A number of studies of stellar maser sources have been made 
toward the central part of the Galaxy (\cite{lin92a}; \cite{izu95}; 
\cite{sjo98}; \cite{deg00a}; \cite{deg04a}). 
This paper particularly focuses on the area of the inner bar region
($|l|<3.5^{\circ}$, and $|b|<1^{\circ}$). Because the IRAS survey was 
very incomplete in this region due to source confusion in the high source
density area, the previous SiO survey based on the IRAS catalog in this region 
(\cite{deg00a} ; $|l|<3^{\circ}$, and $|b|<3^{\circ}$) 
sampled the objects at more or less higher latitudes  ($|b|\gtrsim 1^{\circ}$).
Furthermore, the other unbiased surveys (non-targeted surveys;
\cite{shi97}; \cite{miy01}; \cite{deg02}) 
inevitably focused on the much smaller areas near the galactic center.  

This paper presents the result of SiO maser survey of MSX/2MASS objects
in the area $|l|<3.5^{\circ}$, and $|b|<1^{\circ}$. The source sampling
was made based on the MSX and 2MASS catalogs: the MSX catalog (\cite{pri01}) provides a large number 
($\gg$ 1000) of middle-infrared sources in the above area, except
very near to the galactic center ($\lesssim 10'$). With 2MASS identifications,
the AGB stars under mass loss are easily picked up effectively 
excluding young stellar objects from the sample.
These objects are used as dynamical tracers in the inner bar region of the Galaxy. 
Because the SiO survey of the very central area ($|l|$ and $|b| \lesssim 15'$; inner $\sim$30 pc)
have already finished (\cite{ima02}; \cite{deg04a}), we excluded the central $15'$
area from the present survey.   
   

\section{Observations and results}

Simultaneous observations in the SiO $J=1$--0, $v=1$ and 2 transitions at
43.122 and 42.821 GHz, respectively, were made with the 45-m radio telescope
at Nobeyama during the periods of 2004 February--May, and 2005 February--May.
Prior to the main long-term project of 2004--2005,
a pilot study of about ten sources for this project was made 
in May 2003; detections in this pilot study were also involved in this paper. 
We used a cooled SIS mixer receiver (S40) for the 43 GHz
observations and accousto-optical spectrometer arrays, AOS-H and AOS-W,
having bandwidths of 40 and 250 MHz with 2048 channels each; 
the effective velocity resolution of
the AOS-H spectrometer was 0.3 km s$^{-1}$. They covered the velocity range of
$\pm 390 $ km s$^{-1}$, for both the SiO $J=1$--0 $v=1$ and 2 transitions.
The overall system temperature was between 200 and 300 K,
depending on the weather condition. The half-power telescope beam width
(HPBW) was about 40$''$. The antenna temperature given in the present paper
is that corrected for the atmospheric and telescope ohmic loss, 
but not for the beam or aperture efficiency ($\equiv T_{\rm a}^*$).
The conversion factor of the antenna temperature to the flux density is 
about 2.9 Jy K$^{-1}$. To save observation time, we employed a position-switching 
sequence, Off--On1--On2--On3 to observe three objects at once, 
where the off position was taken 7$'$ west of the first-object position 
(On1) in right ascension; the
separation of the off position corresponds to the angle moved by an object
in the sky during the typical integration (20 s), additional
telescope-slewing, and settling time (10 s), so that the integrations were
made nearly at the same elevation angle.  With this sequence, we saved about 50\% 
of the total observation time compared with the time of simple On-Off sequences. 
Further details of SiO maser observations using the NRO 45-m telescope have been
described elsewhere (\cite{deg00a}), and are not repeated here.

The source selection was made from the MSX and 2MASS catalog, 
by applying the following criteria:

\begin{enumerate}
\item
 $|l|<3.5^{\circ}$ and $|b|<1^{\circ}$.
\item
 MSX objects brighter than $F_{\rm C}=1.75$ Jy (with $F_E>0$)
with 2MASS NIR counterparts within 8$''$, where $F_{\rm C}$ and $F_{\rm E}$ are
the flux densities in the MSX C (12$\mu$m) and E (21$\mu$m) bands, respectively.
\item
 $K<11.5$,  $H-K>1.0$, and $K< 6.0 + 1.6 \times (H-K) $ for the 2MASS counterparts.

\end{enumerate}

The similar criteria except (1) have been used for the previous SiO maser searches 
in the Galactic disk (\cite{nak03}; \cite{deg04b}) and are approved to be quite effective
for picking up the mass-losing AGB stars with SiO masers, though a small number of young 
stellar objects, which have similar colors, inevitably contaminate the sample.
From the sample, we excluded several objects which were already surveyed 
in the same SiO lines (listed in Table 4). We included these objects 
for the statistical analysis in section 3 
(except J17512667$-$2825371=IRAS17482$-$2824, which is apparently a foreground object), 
but not involved in this section.   

We also observed a few, additional interesting objects in this area 
for the sake of completeness; they did not necessarily satisfy 
the above criteria. These are known OH 1612 MHz sources, bright NIR objects,
and bright MSX objects with dubious NIR counterparts.  
  
In total, we observed 291 objects in the area of $7^{\circ} \times 2^{\circ}$
of the Galactic center, resulting in 163 detections in the SiO $J=1$--0 $v=1$ or 2 transitions.
The observational results are summarized in Tables 1 (detections) and 2 (nondetections).
Because the positional accuracy of objects in the 2MASS catalog ($\sim$0.1$''$) 
is better than that of MSX sources (a few arc second), we used
the 2MASS positions for all of the observed objects except for a few sources (which were 
observed in the 2003 pilot study). Therefore, we used the 2MASS conventional names 
for object designation. 
Table 3 summarizes the infrared properties of the observed sources,
listing the 2MASS designation, MSX(6C) designation, separation between 
2MASS and MSX objects, 2MASS $K$ magnitude, 2MASS $J-H$, 2MASS $H-K$, MSX flux density in band C
(12 $\mu$m), and MSX color [$=log(F_{\rm E}/F_{\rm C})$], the radial velocity of SiO, 
the nearest IRAS source within 60$''$, and the separation between the 2MASS and IRAS sources.   
The details of the spectral features and individual objects are
given in Appendix.  

The distribution of the sources in the selected area is shown in figure 1.  
Though the source selection was not biased in the galactic latitude $b$, 
the detections (filled circles in figure 1) were found more 
in the upper half of the $l$-$b$ plane. This is probably because dense clouds
exist more in the $b<0$ side. The $5^{\circ} \times 2^{\circ}$ combined atlas of 2MASS 
and MSX images\footnote{available at 
http://www.ipac.caltech.edu/2mass/gallery /2mass\_msx\_gcatlas.jpg} 
gives several silhouettes against background dust emission at the positions 
coinciding with the source deficient regions in figure 1 around 
$(l,b)$=($-2.3^{\circ}$, $-0.7^{\circ}$), ($-1.7^{\circ}$, $-0.2^{\circ}$), 
($-0.7^{\circ}$,$0.6^{\circ}$), ($0.2^{\circ}$, $-0.6^{\circ}$), ($0.8^{\circ}$, $-0.5^{\circ}$), 
and ($1.7^{\circ}$, $-0.3^{\circ}$); in these area,
source densities in the present sample show an apparent decrease (see figure 1).  
The clouds seen in silhouettes have been considered to be cold dense
clouds without massive star formation (\cite{car98}). Though mapping of dense molecular-clouds
toward the galactic center,  for example, in CS lines does not indicate very large asymmetry 
with respect to the galactic plane (see figure 2a of \cite{tsu99}), the 
distribution of cold dense clouds, which was revealed by the MSX survey, seems apparently 
more abundant in the $b<0$ side. Furthermore the 2MASS atlas, which was noted above, 
does not seem to show strong asymmetry in $b$.\footnote{
see http://www.ipac.caltech.edu/2mass/gallery /2mass\_msx\_gc.html} 
However, this is probably for the 
background level adjustment in making the atlas.    
  
Figure 2 exhibits the NIR magnitude--color (left) and two-color (right) diagrams of
the observed objects. In the two-color diagram, 
objects falling in the area $(J-H)\lesssim (H-K)$
are normally not O-rich AGB stars. However, because of the blending of stars 
due to the high star density toward the Galactic center, 2MASS photometry
does not necessarily give accurate magnitudes. Therefore,
we did not exclude the objects falling in this "forbidden" area 
(a lower right part of the right panel of figure 2) from the sample;
in fact, we detected many objects in SiO maser lines; this effect was already studied by 
\citet{deg04b}.

Figure 3 exhibits the MIR flux-density--color (left) and two-color (right) diagrams of the observed sources. 
Here, the MIR colors are defined as $C_{\rm AC}=log(F_{\rm C}/F_{\rm A})$ and   
$C_{\rm CE}=log(F_{\rm E}/F_{\rm C})$, where $F_{\rm A}$, $F_{\rm C}$ and $F_{\rm E}$ are
MSX flux densities at 8, 12 and 21 $\mu$m (bands A, C, and E), respectively.
Because the IRAS survey was incomplete in this sky area due to source confusion, 
and because the MSX survey sensitivities are considerably different in each band 
(band A is most sensitive), it is interesting to check whether the MIR color--detection rate
relations still keep for the present MSX-based sample.
The right panel of Figure 3 indicates that the SiO detected sources are peaked 
around ($C_{\rm CE}$, $C_{\rm AE}$)=($-0.05$, 0.14), corresponding to the blackbody temperatures
of about 360 and  400 K, which seems slightly higher than the temperature 
obtained from the SiO detection rate of IRAS sources ($\sim 300$ K 
for $C_{12}\simeq 0$ [=$log(F_{25}/F_{12})]$; \cite{izu95}).
Figure 4 exhibits a histogram of the MIR color, $C_{\rm CE}$, 
and a line graph of SiO detection rate.
The detection rate seems to be flat (at about 60\%) between $-0.4$ -- 0.2 in $C_{\rm CE}$,
and drops at both edges, reminding the same tendency of the SiO detection rates
which have been found already by observations of IRAS sources (\cite{izu95}; \cite{deg00a}).  
These graphs assure that the sampling for SiO search was made correctly,
aiming objects with a maximum SiO detection rate.

The extinction-corrected $K$ magnitude (both for interstellar and circumstellar extinctions) 
for an individual object can be calculated from the 
corrected K-band magnitude, 
\begin{equation}
K_{H-K} = K - [A_K/E(H-K)] [(H-K) - (H-K)_0],
\end{equation}
where $A_K$ and $E(H-K)$ are the extinction in the $K$ band (both in interstellar and circumstellar)
and the difference of extinctions at $H$ and $K$ bands (\cite{whi91}). \citet{nis05} gave  $A_K/E(H-K)=1.4$
for NIR objects toward the Galactic Center.
Therefore we use this value for the calculation of the corrected K-band magnitude later on.
We also assume that $(H-K)_0$=0.5, corresponding to the M6III star without interstellar 
and circumstellar extinction (\cite{zom90}).  
If the absolute bolometric magnitude is constant for
the sampled objects, the corrected $K$ magnitude, $K_{H-K}$, can be regarded as an indicator of
the distance. The histogram of $K_{H-K}$ of the present sample is shown in figure 5 
with the SiO detection rate (line graph). The peak of the histogram occurs at $K=5.5$ -- 6.0. 
Note that the M6III star without extinction should have $K\sim 5.5$ at the distance of 8 kpc.  
Considering these facts, we believe that a majority of the sampled objects 
are located at the distances similar to that of the galactic center,
i.e., in the Galactic bulge. Of course, a few bright objects with $K_{H-K}<4$,
or $F_{C}>12$ Jy are probably foreground objects: they were excluded from the sample
in the analysis of the velocity field in the next section. 

We checked the previous OH 1612 MHz or H$_2$O 22.235 GHz masing objects in the same area.
Table 5 summarizes the results: 14 objects (these are mostly detected by the 
VLA or ATCA observations) 
are found to lie within a few arc seconds from the sampled objects, 
except OH357.77$-$00.15, which was a single-dish detection (\cite{cas81}). 
The radial velocities of the OH 1612 MHz objects coincide well with those of
SiO except one case, OH358.720$-$00.620, which was a single peak detection
(\cite{sev97}) of receding side of the expanding shell.


\begin{figure*}
  \begin{center}
    \FigureFile(160mm,60mm){}
  \end{center}
  \caption{Source distribution in the galactic coordinates. The filled and unfilled circles
  indicate the SiO detections and nondetections.The inclined central square indicates 
  the region already surveyed by \citet{deg04a}.
}\label{fig:l-b map}
\end{figure*}

\begin{figure*}
  \begin{center}
    \FigureFile(120mm,80mm){}
  \end{center}
  \caption{Near-IR magnitude-color (left panel) and two-color (right panel) diagrams.
  The filled and unfilled circles
  indicate the SiO detections and nondetections.
}\label{fig:NIR-prop}
\end{figure*}


\begin{figure*}
  \begin{center}
    \FigureFile(120mm,80mm){}
  \end{center}
  \caption{Middle-IR flux-color (left panel) and two-color (right panel) diagrams.
  The filled and unfilled circles
  indicate the SiO detections and nondetections.
}\label{fig:MIR-dia}
\end{figure*}

\begin{figure*}
  \begin{center}
    \FigureFile(90mm,60mm){}
  \end{center}
  \caption{Histogram of the MIR-color, $C_{\rm CE}=log(F_{\rm E}/F_{\rm C})$. 
  The line graph shows the SiO detection rate (with probable uncertainty).
  The shaded and unshaded areas 
  indicate the SiO detections and nondetections.
}\label{fig:detec-rate}
\end{figure*}


\begin{figure*}
  \begin{center}
    \FigureFile(90mm,50mm){}
  \end{center}
  \caption{Histogram of corrected $K$ magnitude, 
  and the SiO detection rate (line graph with uncertainty). 
  The shaded and unshaded areas indicate the SiO detections and nondetections. 
}\label{fig:Kcorr}
\end{figure*}
\begin{figure*}
  \begin{center}
    \FigureFile(100mm,200mm){}
  \end{center}
  \caption{a. SiO longitude-velocity diagram overlaid on the CO $l$--$v$ map
  (taken from \cite{dam01}). Note that SiO represents a stellar motion, while CO
  represents a gas motion. The large circle indicates a foreground object. 
  The 240 pc molecular gas ring feature is faintly seen in the CO map, tracing a parallelogram through the points,
  (2$^{\circ}$, 220 km s$^{-1}$), ($-1^{\circ}$, 100 km s$^{-1}$), ($-2^{\circ}$, $-220$ km s$^{-1}$),
  and (1$^{\circ}$, $-100$ km s$^{-1}$) (cf. figure 6b).
}\label{fig:l-v}
\end{figure*}
\setcounter{figure}{5}
\begin{figure*}
  \begin{center}
    \FigureFile(100mm,200mm){}
  \end{center}
  \caption{b. SiO longitude-velocity diagram overlaid on the $x_1$ and $x_2$ orbit trajectories
[taken from figure 11 of \citet{bis03}].  The large circle indicates a foreground object.
Thin outer curves spreading between $V_{\rm lsr}=\pm 240$ km s$^{-1}$ are trajectories for the $x_1$ orbit family,
and inner curves concentrating between $V_{\rm lsr}=\pm 85$ km s$^{-1}$ and between $l=\pm 1.2^{\circ}$
are those for $x_2$ orbit family.
}\label{fig:l-v2}
\end{figure*}
\begin{figure*}
  \begin{center}
    \FigureFile(60mm,100mm){}
  \end{center}
  \caption{a. longitude--velocity diagram for the eccentric objects (left panel). 
  Filled circle and open square indicate SiO- and OH-detected objects, respectively.
  The blue segments of lines in the left panel are parts of a trial trajectory deducing the spatial orbits. 
  Two objects (filled circle with open square) are of both SiO and OH detections.
}\label{fig:high_vel}
\end{figure*}
\setcounter{figure}{6}
\begin{figure*}
  \begin{center}
    \FigureFile(60mm,100mm){}
  \end{center}
  \caption{b. longitude--velocity diagram for the inner-bar objects with SiO masers.  
  Open and filled circles indicate the suggested $x_1$ and $x_2$ orbit objects. 
  The red solid lines indicate the boundaries of subsets 
  for the orbital families, and the red broken line
  indicates a separation for the fore and back sides of the $x_2$ orbits
  (see the right panels of figure 10).
}\label{fig:inner-bar}
\end{figure*}

\begin{figure*}
  \begin{center}
    \FigureFile(80mm,50mm){}
  \end{center}
  \caption{Histogram of $K_{H-K}$ for the eccentric-group objects. 
  The shaded and unshaded areas indicate the $V_{\rm lsr}>0$ 
  and $V_{\rm lsr}<0$ sets.
}\label{fig:Hist-out}
\end{figure*}


\section{Discussion}

\subsection{Longitude--velocity diagram : overall structure}

The radial velocities of the 163 SiO detected objects with additional 5 previous detections
in table 4 (except IRAS 17482$-$2824 for a probable foreground object) are plotted 
against Galactic longitude in figure 6. The CO $J=1$--0 $l$--$v$ map 
(\cite{dam01}) and the map of the $x_1$ and $x_2$ orbits in a bar potential (\cite{bis03}) 
are overlaid in figures 6a and 6b, respectively. 
The SiO and CO distributions seem to be moderately correlated:
the source-sparse regions (we often called these as "holes" in previous papers; \cite{izu95}) 
are seen around ($l$, $V_{\rm lsr}$)=($-2.4^{\circ}$, 30 km s$^{-1}$), and
($-2.2^{\circ}$, $-120$ km s$^{-1}$), 
where the CO map (which indicates gas distribution toward the same direction) also
gives a low emission intensity. More SiO sparse regions are seen 
around (1.5$^{\circ}$, 50 km s$^{-1}$) and ($-0.5^{\circ}$, 30 km s$^{-1}$),
which coincide with the vacant regions of the $x_1$ and  $x_2$ trajectories in figure 6b.
These features are considered to be observational signatures of bar dynamics (\cite{kui95}).
However, a large difference between the SiO and CO distributions 
is recognized at the extreme high velocity part, i.e.,
outer envelope of the CO features; SiO objects are seen beyond the upper and lower limits of
the radial velocity of the CO gas, which will be discussed in the next section.  
In addition, no strong feature in the SiO distribution is found for the velocities
corresponding  the 240 pc molecular ring (cf. \cite{bin91,saw04}).

Furthermore, figure 6a indicates that more SiO objects appear at the negative
velocity side than at the positive velocity side (113 against 55,
respectively).  This asymmetry of the source velocity distribution
has already been noticed in \citet{izu95} and \citet{deg00a}; this phenomenon
has been seen in all of the SiO maser velocity data toward the Galactic bulge.
This is attributed to the preferential detection of the bulge foreside objects:
the foreside (to the Galactic center) stars tend to have negative radial velocities
and backside stars tend to have positive radial velocities due to bulge
streaming motion in the $x_1$ orbits; preferential detections of the foreside sources
with the telescope sensitivity reaching to the Galactic center objects
make more objects with the negative velocities.  

The other notable feature in figure 6 is a rough alignment of the objects 
along the line $V_{\rm lsr}\sim 100 \times (l/{\rm deg})$ km s$^{-1}$
(which is shown as the red broken line in figure 7b)
at the range  $|l|<1.5^{\circ}$, which roughly coincides with
 the $x_2$ orbit trajectories. This feature was not seen
in the $l$--$v$ map of IRAS/SiO objects in $|l|<3^{\circ}$ and $|b|<3^{\circ}$
(figure 6 of \cite{deg00a}),
in which the survey based on IRAS catalog was quite incomplete in $|b|<1^{\circ}$.
Therefore, it is highly likely that these are objects in the $x_2$ orbits.
A full discussion of these objects will be given in Section 3.3.

Figures 7ab illustrate the high-velocity group of stars outside of the $x_1$ orbits
and the low velocity group which presumably belongs to the $x_2$ orbit family.
The boundary was drawn somewhat arbitrarily, though the source distributions and
the model orbits were taken into account.
        
\subsection{High-velocity objects in the longitude--velocity diagram}

Figure 7a depicts the eccentric-group objects in the $l$--$v$ diagram.
These objects are chosen because of their positions out of $x_1$ orbit trajectories. 
They cannot belong to the normal outer $x_1$ orbits, 
because the $x_1$ orbits with larger apocenter
distances take smaller $|V_{\rm lsr}|$ in the $l$--$v$ diagram,
where the trajectories of such orbits are extended to the middle-right 
to middle-left (with an inclination due to a rotational motion) as shown in figure 6b.

Note that the high-velocity objects 
were found at very inside ($|l|\sim$ a few arcminutes from the Galactic center; \cite{van92,deg04a})
to the outside, i.e., $|l|\sim 3^{\circ}$ out of 
the molecular ring at $R=\sim 240$ pc.
They are presumably
in the orbits with peri- and apo-center distances of $\sim 0.01$ and $\gtrsim 1.5$ kpc,
respectively, gaining a large kinetic energy with converting the potential energy. 
Because a stellar system is collisionless,
their orbits can be intersected, even though such orbits are forbidden in the gas dynamics 
(e.g., \cite{bin91}); the intersection of orbits causes strong shock in gas dynamics, 
quickly infalling gases to the inside (\cite{con89}).    

An inferred trajectory of the eccentric group in the $l$--$v$ diagram (figure 7a),
which is suggested from the SiO observation in the present paper,  
is shown in thick blue lines.
We suggest that these are the objects in highly eccentric orbits along the bar
with higher speed than normal $x_1$ orbit family
(as shown in the lower left panel in figure 10). 
In this model, the orbit is prograde; the
object in front of the galactic center should exhibits
a large negative radial velocity and the object behind should show a
large positive velocity. 

In order to check the above hypothesis,
we have calculated the average extinction-corrected $K$ magnitude, $K_{H-K}$,
for the eccentric group in present SiO sample. The statistical analyses are
summarized in Table 6.
We divided the 22 SiO eccentric objects into two groups: one with $V_{\rm lsr}<0$ and
the other with $V_{\rm lsr}>0$ (upper and lower parts in figure 7a).  
Figure 8 shows a histogram of $K_{H-K}$ for each group.
We obtain [$K_{H-K}$]$_{\rm ave}$=5.7 and 6.0 for the $V_{\rm lsr}<0$ and $>0$ groups, respectively. 
Therefore, we again confirmed the tendency of the negative-velocity objects having 
larger brightness, though it is not statistically significant because of the small number;
the t-test gives a 43 \% probability for the null hypothesis of 
two sets having the same distribution function; in other words, only with 57 \% 
of the confidence level, we can state that the average $K_{H-K}$ are different 
between the two sets. 

In order to have higher confidence level in statistics, we have added 9
similar high-velocity eccentric objects in $|l|<3^{\circ}$ and $|b|<3^{\circ}$,
which was found in the previous SiO survey (\cite{deg00a}). The results are
summarized in the middle rows of table 6. With slight increasing numbers of the sets, 
the statistical confidence level increases up to 66 \%.
Furthermore, even if we added the other 15 eccentric objects 
(with 2MASS $K$-band identification), which were found by OH 1612 MHz survey 
in the same area (\cite{sev97}), the statistical confidence does not increase
significantly. In fact, the addition of OH 1612 MHz causes a significant increase of
[$K_{H-K}$]$_{\rm ave}$ for both groups, as shown in table 6. Therefore, the NIR property of
the OH 1612 MHz objects (detected by \cite{sev97}) is inferred considerably different 
from the SiO masing objects, probably due to contamination by post-AGB objects 
(or even by young stellar objects) to the sample. 
In fact, 2/3 of the sampled OH 
high-velocity objects are at higher latitudes ($1^{\circ}<|b|<3 ^{\circ}$),
suggesting that high velocity objects appear at considerably higher latitudes
in the Galactic bulge (even higher than $|b|>3^{\circ}$ as suggested by
higher-latitude SiO maser surveys; \cite{izu95}).  

Above analyses revealed the tendency of the negative (or positive) high velocity objects
being in front of (or behind) the Galactic center,  
although it is not statistically very significant. 
One reason for obtaining this 
indecisive result seems to stem from the use of 2MASS $K$ magnitudes which were
given only in one-time measurements. Because SiO masing objects are
mostly large-amplitude variables as miras and semiregulars (for example, see \cite{deg04a}), 
the average of $K_{H-K}$ in the present sample inevitably involve uncertainty
of about 1 magnitude (see \cite{gla01}). We believe that, if the $K$ magnitude averaged over a 
light variation period is used, a significant improvement will be obtained in the statistical analysis.
The other factor of uncertainty may be involved in the 2MASS photometric measurement
due to blending of the star images. 
However, because of the small number,
exclusion of these low-quality objects does not 
improve the situation in the present data.

The velocity distribution of stars in the bar-like bulge can be interpreted in terms
of the periodic orbits which compose the bar potential (\cite{sch79}).
A number of theoretical studies of the stellar orbits in the barred spiral 
galaxies revealed that there are stable periodic orbits called 
as the $x_1$ and $x_2$ orbit families (\cite{con88,ath92}) with additional higher-order resonance family 
and the other stationary-orbit family in which a particle moves around the periodic orbits. 
(\cite{bur99,bur05}). Dynamical models for the Galactic bulge
have been built with Schwartzschild's galaxy building technique (\cite{zha96}; \cite{haf00}; \cite{bis04}).
However, a rigorous comparison of the model velocity fields or distribution function with
observational ones has still been lacked. According to the model created
by \citet{zha96}, 55\% of stars in the bulge belong to the direct regular (mostly 
$x_1$) orbits, and the other 45\% of stars are in composite 
(irregular and retrograde) orbits. The dynamical model created by \citet{haf00}
suggested a slight enhancement of mass distribution of stars at near-zero angular momenta
with considerable high energies (figure 12 of \cite{haf00}), though exact percentage of
these stars and their spatial orbits were not given explicitly. It is desirable
to compare the fraction of these low angular-momentum objects in these numerical models 
with the present result in future. More discussion on this issue is given in Appendix B.

\subsection{Inner-bar objects in the longitude--velocity diagram}

As described in section 3.1, we find a rough alignment of stars to the straight line (near the
$x_2$-family orbits ; figure 6b). 
A similar feature was seen in the $l$--$v$ diagram of OH 1612 MHz data subset 
($|l|< 1^{\circ}$ and $|b|< 0.5^{\circ}$ ; \cite{sev97}), which was shown 
clearly in figure 8 of \cite{deg00a} in filled symbols. 
The feature seems to appear less clearly 
in the lower right quadrant of figure 6b,
because the $x_1$-family objects are expected to contaminate the sample more 
in the negative radial velocity than in the positive radial velocity.

We depicted the part of $l$--$v$ diagram in figure 7b,
prescribing the boundaries of the $x_2$ orbit feature based on the following considerations. 
Because of the concentration 
of 6 objects at the upper area, $0^{\circ}<l<1.5^{\circ}$ and $100 <V_{\rm lsr}< 150$ 
km s$^{-1}$, we designated these also in the $x_2$ family, though
the theoretical curves for the $x_2$ family shown in figure 6b (\cite{bis03}) reach only 
up to $V_{\rm lsr}=80$ km s$^{-1}$. 
In fact, the $x_2$ orbits in the models of \citet{eng99} extend up to $\pm 120$--140 km s$^{-1}$
in their figures 13 and 14. The difference from \citet{bis03} seems to stem from the rapid rise of
rotation curve at the inner 300 pc in the \citet{eng99}'s model. 
Therefore,  a considerable freedom seems to exist
in dynamical models for the limiting stellar velocities in $x_2$ orbits,
though some constraints can be obtained from CO gas observations (e.g., \cite{sta04}).
Furthermore, stellar (composite) orbits can possibly be extended beyond
the outer boundary of gas $x_2$ orbits.   

Numerical model calculations predicted a very low fraction (less than 1 \%) of 
$x_2$ orbits stars among the whole bulge stars (\cite{zha96,haf00}). However,
the present SiO observation seems to give more than a few percent of stars 
belong to the $x_2$ orbit family, though exact evaluation of the fraction is
somewhat difficult due to lower survey completeness of SiO maser sources 
at the outer part of the bulge. Moreover,    
the boundary drawing in figure 7b is slightly subjective. 
Therefore, we made a statistical analysis for the designated object sets 
and checked the validity of grouping
in figure 7b. 

We performed the following two statistical tests.
The first check is the latitude distribution for the two sets:
one for the supposed $x_2$ family set (filled circles in figure 7b), 
and the other for the supposed $x_1$ family set (open circles in figure 7b).
It is expected that the $x_2$ family orbits are centrally condensed inside the 
$x_1$ orbits. Therefore, the $x_2$ family objects spread less in Galactic latitude. 
The first two rows of table 7 summarize the result of statistical analysis:
the average of $|b|$ for the supposed $x_2$ family set  is $0.40^{\circ}$ ($\pm 0.31^{\circ}$)
while that for the supposed $x_1$ family set is $0.53^{\circ}$ ($\pm 0.27^{\circ}$).
The t-test gives the probability of 0.10 for the null hypothesis of two sets 
having identical distribution functions. In the other words, we can state 
with a probability of more than 90\% that the selected $x_2$ family set is more
concentrated to the galactic plane than the selected $x_1$ family set is.

Above analysis indicates that the selection of the $x_2$ family objects
is made roughly correctly, though the $x_1$ family objects may contaminate
the $x_2$ family set in some degree (the opposite is less likely). 
Therefore, we investigated further whether or not
evidence appears on the depth difference along the line of sight for these objects.  
We separated the sample into 4 sets: the foreground $x_1$-family set, 
the background $x_2$-family set, the foreground $x_2$-family set, 
and the background $x_1$-family set (most left, next left, next right, most right
in figure 7b [the $x_2$-family set devided into two by the broken line]; 
see the spatial geometry in the lower right panel of figure 10).  
The result is shown after 
the 3rd raw of Table 7. In these sets, set 1 has brightest $K_{H-K}$,
and set 2 has faintest $K_{H-K}$. The t-test gives that the difference
of $K_{H-K}$ between set 1 and set 2 is statistically significant with the 90\% significance level, 
though not for other pairs of the sets.  

In this analysis, it is robust that there is a significant difference of about $\sim 0.5$ mag 
in mean $K_{H-K}$ mag for the foreside $x_1$ family objects from others. 
This value corresponds to the distance ratio of about a factor of 1.25,
indicating that the set-1 objects are located at distance of 6.4 kpc from the Sun on the average 
if we assume the distance of 8 kpc for the other sets (average of sets 2 and 3) of objects. 
It is not clear why the average $K_{H-K}$ of the set 4 is brighter than the average $K_{H-K}$
of set 2. It may be due to statistical fluctuation by the small number of set 4, or contamination by members 
in composite-orbit families. 

\section{Conclusion}
We surveyed the MSX/2MASS objects in the sky area of $|l|<3.5^{\circ}$ and $|b|<1^{\circ}$
in the 43 GHz SiO maser lines and obtained accurate radial velocities of the detected 163 objects. 
The $l$--$v$ diagram of this sample clearly revealed the $x_2$ orbit family feature 
at the positive velocity side. Furthermore, an eccentric velocity feature,
which cannot be attributed to the $x_1$ orbit family, appears
on the same diagram. The extinction-corrected $K$ magnitude (a distance modulus 
when subtracted by absolute magnitude),
indicates a sequential deposition of these feature sets if we properly
assign the fore and back sides of the bulge-bar objects from their radial velocities.
Though the statistical tests do not necessarily give  
significant results except a few cases, the tendency appeared in
the distance modulus is consistent with the dynamical model
using periodic orbits in the bulge bar. We conclude that the measurements of average 
$K$ magnitudes over light-variation periods will be useful for further accurate 
determination of the distance, as well as the presentation of the 
periodic orbits on the position--velocity diagram will be for comparison with the observations.

\

The authors thank Dr. T. Omodaka, and his students for the help 
of observations and encouragements. They also thank Drs. T. Sawada and J. Koda
for useful discussions.
This research made use of the SIMBAD and VizieR databases operated at CDS, 
Strasbourg, France, and as well as use of data products from 
Two Micron All Sky Survey, which is a joint
project of the University of Massachusetts and Infrared Processing 
and Analysis Center/California Institute of Technology, 
funded by the National Aeronautics and Space Administration and
National Science foundation, and from the Midcourse Space 
Experiment at NASA/ IPAC Infrared Science Archive, which is operated by the 
Jet Propulsion Laboratory, California Institute of Technology, 
under contract with the National Aeronautics and Space 
Administration.   

\section*{Appendix. A. Individual objects}

We presented all of the spectra of the SiO $J=1$--0 $v=1$ and 2 transitions
for the detected objects in figure 9a--j. Here we discuss the individually
interesting objects. 

\begin{itemize}
\item
J17404953$-$3055183 ($\sim$OH 357.77$-$00.15 =IRAS 17375$-$3053):
Doubly peaked OH 1612 MHz object was detected by Parkes 64m telescope ($HPBW \sim 12.5'$ beam)
about 37$''$ North-East of this object (\cite{cas81}), and the position is close
to the faint MSX object, G357.7672$-$00.1451 ($F_{\rm C}=0.8$ Jy), 
and IRAS 17375$-$3053. However, slightly
brighter MSX object with $F_{\rm C}=2.4$ Jy, G357.7643$-$00.1591, 
is the object that we observed in the present paper. 
Because \citet{sev97} did not detect this object with VLA, 
we do not have accurate positions of this OH source.
Because the radial velocity of SiO 
source differs only by 5.8 km s$^{-1}$ to the center velocity of OH double peaks,
we tentatively assigned this object to OH 357.77$-$00.15 in table 5.

\item
J17425115$-$2951513 (=G358.8932+00.0277):
This is a bright infrared object with $F_C=27.2$ Jy
with no IRAS counterpart within 1$'$. 
The SiO $J=1$--0 $v=1$ spectrum exhibits interesting triple peaks
spreading in the velocity span of about 20 km s$^{-1}$, which may be an indication
of a supergiant with heavy mass loss. No previous OH maser observation was 
reported for this source.

\item
J17465905$-$2817037? (=ISOGAL$-$P J174659.1$-$281658=OH .713 +.084):
This is a faint OH/IR object which occupies a unique position 
in the left panel of figure 3 ($K=11.193$ and $H-K=1.715$),
where it is out of the area aimed in the present SiO survey. 
This object was observed because of the previous OH detection
 (\cite{lin92a}) and its peculiar characteristic in the $K$--$H-K$ diagram. 
The SiO radial velocity, 98.4 km s$^{-1}$, agrees well with
the center velocity of OH double-peaks,  96.8 km s$^{-1}$.

\item
J17481382$-$2725523 (=G001.5844+00.2889 =IRAS 17450$-$2724 ) : 
This is a bright IR source
with $F_{\rm C}=44.8$ Jy) and IRAS LRS class of 13 (no clear spectral feature). 
This object was observed for the sake of completeness in this SiO survey.
SiO maser emission was detected first time in this survey 
at $V_{lsr}=-29.5$ km s$^{-1}$, identifying this object 
as an O-rich evolved  star. No OH 1612 MHz line has ever been detected.

\item
J17520868$-$2711412 (=G002.2335$-$00.3360 =IRAS 17490$-$2711):
This is a bright IR source with $F_{\rm C}=15.7$ Jy). 
SiO maser emission was detected first time in this survey 
at $V_{lsr}=-81.5$ km s$^{-1}$, identifying this object 
as an O-rich evolved star.

\end{itemize}

\section*{Appendix. B. Approximate Orbits}
From the deduced trajectory in the $l$--$v$ diagram (figures 7ab), we can draw
an approximate orbital curve on the 2D coordinate system corotating with a bar. 
Here we assume that a pattern velocity of the bar, $\Omega _p$=60 km s$^{-1}$ (\cite{bis03}).
The particle position on the line of sight can be written as
\begin{equation}
Y=Y_0 + \int V_Y dt = Y_0 + \int V_Y/V_X dX.
\end{equation}
Here the $Y$ axis is taken along the line of sight toward the galactic center
and the $X$ axis is taken to be perpendicular to the Y axis on the galactic plane, 
and $V_X$, and  $V_y$ are the velocity components 
on the X and Y directions on the co-rotating system, respectively, and $Y_0$ is the integration constant. 
The velocity, $V_Y$, is given from the assumed trajectory ($V_r$)
as a function of the coordinate $X$ in the corotating system as
\begin{equation}
V_Y \sim V_{r} - X \; \Omega _p.
\end{equation}
Here, $V_r$ is the radial velocity of a star in the rest frame ($\equiv V_{lsr} + V_0 \, sin \, l$, where 
$V_0$ and $l$ are the rotational velocity of the Local Standard of Rest and the Galactic longitude,
respectively).
Because the velocity perpendicular to the line of sight, $V_X$, is unknown,
we assume for simplicity that $V_X$ is constant on each segment of the trajectory.
Then, equation (2) indicates that
the linear segment of the trajectory in the $l$--$v$ diagram is expressed by
a part of parabola in the $X$--$Y$ coordinates.
The four linear segments of the trajectory (upper panel of fig 10) can be
drawn as a closed loop of four segments of parabola 
as shown in the lower panel of figure 10.
(Because the straight dotted lines connecting the upper and lower edges in the upper-left panel are 
just a first guess, they are not verified except for the case of the 2:1 resonance.
For the case of 4:1 resonance,
these parts are inferred to be complex crossed curves.)

Because the assumption of constant $V_X$ is not exactly hold
for the realistic cases, an orbit obtained with this method
is quite approximate. However, the obtained looped orbit
correctly shows an elongation along the large scale bar major axis,
and four corners of the orbit suggesting correctly the self-crossing looped orbit,
even though the width of the loop can vary with
arbitrary choice of constant value of $V_X$.      
The obtained loop orbit
is similar to the $x_1$ family orbits
shown in figure 6 of \citet{sko02}, or self-crossing orbits in the figure 13 of \citet{reg04}.
However, because the orbits, which are inferred from the observation, do not satisfy any dynamical
requirements, the stars on these orbits may not necessarily rotate at the bar pattern speed
in a loger time scale. 
The difference of the pattern speed makes
the pericenter precession, and a real orbit is likely to be a non-closing loop.
Such a precessing orbital motion makes the limiting radial velocities
smaller (with making large angles to the line of sight from the Sun), 
as illustrated by the broken lines in the upper-left panel in figure 10 
(for example of a dynamical model, see figures 4--6 of \cite{pic04}).
In the case when the precessing motion is rapid (higher-order resonance or composite orbits), 
the particle density in the $l$--$v$ diagram 
tends to be homogeneous in the broken square in the upper-left panel in figure 10.
With such highly irregular or composite orbits, it may be hard to explain the presence 
of "holes" in the  $l$--$v$ diagram
as shown in figure 6. Therefore, the slow-precession, as for the
4:1 resonance family orbits (as illustrated by \cite{con88,kau05}), 
would be preferable as an explanation of the
eccentric group of stars. 
   
\setcounter{figure}{9}
\begin{figure*}
\vspace{0cm}
  \begin{center}
    \FigureFile(150mm,160mm){}
  \end{center}
  \caption{Schematic trajectory in the position-velocity diagram (upper panels) 
  and the corresponding spatial orbit
  in the corotating system (lower panels). The left panels illustrate 
  the case of eccentric-velocity group and the right panels the case of the $x_2$ orbits. They are assumed
  to be expressed by the segments of linear trajectory in position-velocity diagram, but in fact,
  they must be more smoothed curves. The open and filled arrows show the directions of motion. 
  The position--velocity diagram ($X$--$V_Y$) can be converted to the $l$--$V_{\rm r}$ diagram 
  by adding the rotation of the coordinate, $X \; \Omega_{\rm P}$. Dotted lines in the left panels
  are just guesses connecting the upper and lower segments, which may be more complex curves 
  for 4:1 or higher-order resonace orbits.  Broken lines in the top left panel 
  indicate the upper and lower limits for the orbits precessing
  relatively to the bar pattern, which is indicated by broken arrows in the lower-left panel.
  The Sun is supposed to be located at (0, $-8$) in the lower panels.
  }\label{fig:orbit}
\end{figure*}



\tabcolsep 2pt
\begin{longtable}{lrccc@{\hspace{2mm}}rccc@{\hspace{2mm}}rr}
\caption{SiO detections.}
\hline
%
     & \multicolumn{4}{c}{SiO $J=1$--0 $v=1$} & \multicolumn{4}{c}{SiO $J=1$--0 $v=2$} & & \\
Name & $V_{\rm peak}$ & $T_{\rm peak}$ & Flux & RMS & $V_{\rm peak}$ & $T_{\rm peak}$ & Flux & RMS & Obs. date & $V_{\rm ave}$\\
     & {\tiny (km s$^{-1}$)} & {\tiny (K)} & {\tiny (K km s$^{-1}$)} & {\tiny (K)} 
	 & {\tiny (km s$^{-1}$)} & {\tiny (K)} & {\tiny (K km s$^{-1}$)} & {\tiny (K)} & {\tiny (yymmdd.d)} & {\tiny (km s$^{-1}$)} \\
\hline
\endfirsthead
\hline
     & \multicolumn{4}{c}{SiO $J=1$--0 $v=1$} & \multicolumn{4}{c}{SiO $J=1$--0 $v=2$} & & \\
Name & $V_{\rm peak}$ & $T_{\rm peak}$ & Flux & RMS & $V_{\rm peak}$ & $T_{\rm peak}$ & Flux & RMS & Obs. date & $V_{\rm ave}$\\
     & {\tiny (km s$^{-1}$)} & {\tiny (K)} & {\tiny (K km s$^{-1}$)} & {\tiny (K)} 
	 & {\tiny (km s$^{-1}$)} & {\tiny (K)} & {\tiny (K km s$^{-1}$)} & {\tiny (K)} & {\tiny (yymmdd.d)} & {\tiny (km s$^{-1}$)} \\
\hline
\endhead
\hline
\endfoot
\hline
\multicolumn{8}{l}{$\dagger$ the MSX position was observed. } \\
\endlastfoot
J17352480$-$3058174 & $-$121.2 & 0.348 & 1.034 & 0.045 & $-$121.8 & 0.346 & 0.634 & 0.048 & 040225.3 & $-$121.5 \\
J17354457$-$3134332 & 59.8 & 0.461 & 0.832 & 0.047 & 59.8 & 0.392 & 0.862 & 0.061 & 040422.1 & 59.8 \\
J17355075$-$3057569 & $-$270.1 & 0.306 & 0.658 & 0.047 &  ---  &  ---  &  ---  & 0.048 & 040225.3 & $-$270.1 \\
J17355648$-$3140399 & $-$170.0 & 0.789 & 3.326 & 0.047 & $-$168.3 & 0.714 & 2.925 & 0.054 & 040422.1 & $-$169.1 \\
J17355814$-$3101029 & $-$67.6 & 0.178 & 0.598 & 0.041 & $-$63.2 & 0.227 & 0.500 & 0.054 & 040225.3 & $-$65.4 \\
J17355954$-$3136317 & $-$334.2 & 0.314 & 0.436 & 0.045 & $-$334.5 & 0.271 & 0.748 & 0.057 & 040422.1 & $-$334.3 \\
J17361848$-$3038053 & $-$99.3 & 0.492 & 2.130 & 0.045 & $-$98.2 & 0.451 & 1.906 & 0.059 & 040422.1 & $-$98.7 \\
J17364217$-$3059117 & $-$145.8 & 0.296 & 0.644 & 0.046 & $-$146.0 & 0.318 & 0.935 & 0.054 & 040511.0 & $-$145.9 \\
\hline
\end{longtable}
\begin{longtable}{lrrc}
  \caption{Negative results.} 
\hline  
object name & RMS($v=1$) & RMS($v=2$) & Obsdate \\
          & (K)     &  (K)     & (yymmdd.d) \\
\hline
\endfirsthead
\hline
object name & RMS($v=1$) & RMS($v=2$) & Obsdate \\
          & (K)     &  (K)     & (yymmdd.d) \\
\hline
\endhead
\hline
\endfoot
\hline
\multicolumn{4}{l}{$\dagger$ : the MSX position was observed.}
\endlastfoot
J17352674$-$3109109  &  0.060 & 0.049 &  40511.0 \\
J17360128$-$3114433  &  0.061 & 0.050 &  40511.0 \\
J17372673$-$3132504  &  0.078 & 0.059 &  50527.1 \\
J17375350$-$3135225  &  0.079 & 0.060 &  50527.1 \\
J17381178$-$3146270  &  0.052 & 0.040 &  40421.1 \\
J17381803$-$3053066  &  0.086 & 0.073 &  40421.2 \\
J17382915$-$3106140  &  0.047 & 0.040 &  40421.1 \\
J17383271$-$2946546  &  0.068 & 0.052 &  40515.0 \\
\end{longtable}

\tabcolsep 2pt
\begin{longtable}{llcccccrrcr}
\caption{Infrared properties of the observed sources} 
\hline
2MASS name & MSX6C name & $\Delta r_{M}$ & $K$ & $J-H$ & $H-K$ & $F_C$ & $C_{CE}$ & $V_{\rm lsr}^{SiO}$ & IRAS name & $\Delta r_{I}$ \\
           &            & ($''$)          & (mag) & (mag) & (mag) & (Jy)  &        & {\tiny (km s$^{-1}$)}    &            & ($''$)  \\
\hline 
\endfirsthead
\hline
2MASS name & MSX6C name & $\Delta r_{M}$ & $K$ & $J-H$ & $H-K$ & $F_C$ & $C_{CE}$ & $V_{\rm lsr}^{SiO}$ & IRAS name & $\Delta r_{I}$ \\
           &            & ($''$)          & (mag) & (mag) & (mag) & (Jy)  &        & {\tiny (km s$^{-1}$)}    &            & ($''$)  \\
\hline
\endhead
\hline
\endfoot
\hline
\multicolumn{5}{l}{$\dagger$ : the MSX position was observed.}
\endlastfoot
J17352480$-$3058174  & G357.1017+00.7945  & 0.8 & 8.688 & 2.855 & 1.999 & 2.61 & $-$0.062 & $-$121.5 & 17321$-$3056  & 6.4 \\
J17352674$-$3109109  & G356.9526+00.6906  & 0.7 & 8.823 & 2.864 & 1.758 & 2.94 & $-$0.049 &  &  &  \\
J17354457$-$3134332  & G356.6311+00.4090  & 0.7 & 7.734 & 3.632 & 2.429 & 9.57 & $-$0.091 & 59.8 &  &  \\
J17355075$-$3057569  & G357.1567+00.7196  & 0.9 & 8.204 & 3.314 & 1.916 & 2.29 & 0.001 & $-$270.1 & 17326$-$3056  & 49.1 \\
J17355648$-$3140399  & G356.5680+00.3184  & 0.7 & 11.077 & 0.976 & 4.417 & 2.67 & $-$0.138 & $-$169.1 & 17327$-$3138  & 23.3 \\
J17355814$-$3101029  & G357.1272+00.6694  & 0.8 & 6.791 & 2.668 & 1.632 & 4.34 & $-$0.162 & $-$65.4 &  &  \\
J17355954$-$3136317  & G356.6319+00.3463  & 0.4 & 8.778 & 3.682 & 3.200 & 4.93 & 0.084 & $-$334.3 &  &  \\
J17360128$-$3114433  & G356.9413+00.5373  & 1.3 & 7.820 & 3.340 & 1.984 & 2.28 & $-$0.107 &  & 17327$-$3112  & 2.6 \\
\end{longtable}
\tabcolsep 2pt
\begin{longtable}{lllcrrrrrrr}
\multicolumn{11}{c}{Table 4. Previously observed sources (detected in \cite{deg00a}).}\\
\hline\hline
2MASS name & MSX6C name & $\Delta r_{\rm M}$ & $K$ & $J-H$ & $H-K$ & $F_{\rm C}$ & $C_{\rm CE}$ & $V_{\rm lsr}^{\rm SiO}$ & IRAS name & $\Delta r_{\rm I}$ \\
           &            & ($''$)          & (mag) & (mag) & (mag) & (Jy)  &        & {\tiny (km s$^{-1}$)}   &            & ($''$)  \\
\hline 
J17371631$-$3131192  & G356.8515+00.1635  & 1.8 & 8.266 & 6.134 & 3.704 & 8.24 & $-$0.139 & $-$85.9 & 17340$-$3129  & 25.7 \\
J17391263$-$3044163  & G357.7361+00.2321  & 0.9 & 6.645 & 3.836 & 2.246 & 5.72 & $-$0.105 & --- & 17359$-$3042  & 2.7 \\
J17404994$-$2946001  & G358.7448+00.4516  & 1.3 & 8.650 & 3.826 & 2.347 & 3.21 & $-$0.007 & 0.3 & 17376$-$2944  & 2.4 \\
J17431499$-$2853280  & G359.7669+00.4658  & 0.6 & 7.359 & 3.640 & 1.974 & 2.21 & $-$0.180 & ---$^{\flat}$ & 17400$-$2852  & 17.7 \\
J17494019$-$2844464  & G000.6207$-$00.6604  & 2.5 & 10.147 & 1.809 & 4.832 & 11.81 & $-$0.071 & $-$46.2 & 17464$-$2843  & 16.0 \\
\end{longtable}

\tabcolsep 2pt
\begin{longtable}{llrrrc}
\multicolumn{6}{c}{Table 5. Previous OH and H$_2$O detections}\\
\hline\hline
2MASS name & OH/H$_2$O name & $\Delta r$ &  $V_{\rm lsr}^{\rm OH/H_2O}$ & $V_{\rm lsr}^{\rm SiO}$ & reference \\
           &                & ($''$)     & {\tiny (km s$^{-1}$)}  & {\tiny (km s$^{-1}$)} & \\
\hline 
J17354457$-$3134332   &  OH 356.631+00.409  & 0.5 &  59.7 &  59.8 & (1)  \\
J17404953$-$3055183   &  OH 357.77$-$00.15(?) &37.3 & $-$80.0 & $-$85.8 & (2) \\
J17405280$-$3009137   &  OH 358.422+00.238  & 0.4 & 122(S)&  ---  & (1)  \\  
J17413828$-$3145337   &  OH 357.146$-$00.750  & 0.2 &  21.1 &  21.1 & (1)  \\
J17421887$-$3015288   &  OH 358.497$-$00.081  & 0.7 & $-$22.0 & $-$24.2 & (1)  \\
J17422046$-$2853213   &  OH 359.664+00.636  & 1.0 &$-$160.0 &  ---  & (1)  \\
J17431231$-$2928496   &  H$_2$O 359.260+00.164 & 0.7 &  54.8 &  75.5 & (3) \\ 
J17434496$-$2941008   &  OH 359.149$-$00.043  & 1.3 &  37.1 &  ---  & (1)  \\
J17445891$-$3021064   &  OH 358.720$-$00.620  & 0.4 & $-$20.6(S)& $-$40.0 & (1)  \\
J17451381$-$2941498   &  OH 359.681$-$0.095   & 6.3 & $-$98.7 &  ---  & (4) \\
J17483328$-$2825103   &  OH 000.775$-$00.282  & 0.7 & 192(S)&  ---  & (1)  \\
J17485495$-$2635343   &  OH 002.382+00.590  & 0.1 &  94.8 &  95.7 & (1)  \\
J17502910$-$2550299   &  OH 003.207+00.675  & 0.6 & $-$60.7 & $-$60.4 & (1)  \\
J17531981$-$2646451   &  OH 002.726$-$00.352  & 0.7 & $-$143.2&  ---  & (1)  \\
\hline
\multicolumn{6}{l}{reference: (1) \citet{sev97}, (2) \citet{cas81},} \\
\multicolumn{6}{l}{ (3) \citet{tay93}, (4) \citet{sjo98}} \\
\end{longtable}

\setcounter{table}{5}
{\setlength{\tabcolsep}{2pt}\footnotesize
\begin{longtable}{c|cccc|c}
  \caption{Results of statistical analysis for the eccentric velocity sets}\label{tab6}
  \hline\hline
 Set & group  &  number  & $K_{H-K}^{ave}$ &  stand.dev. & t-probability \\
     &        &          &                 &              &          \\
\endfirsthead
	\hline
 Present SiO sample &  $V_{\rm lsr}<0$  &  15  &   5.727  &  0.881 & 0.43 \\
                    &  $V_{\rm lsr}>0$  &   7  &   5.975  &  0.547 &    \\
  &&&&&\\ 
 added IRAS/SiO objects &  $V_{\rm lsr}<0$  &  19  & 5.871 & 1.009  & 0.34  \\
 (within $|l|<3^{\circ}$ and $|l|<3^{\circ}$)  &  $V_{\rm lsr}<0$  &  12  & 6.197 & 0.839  &   \\
   &&&&&\\ 
 added OH 1612 MHz objects &  $V_{\rm lsr}<0$  & 27 & 6.432 &  1.662 &  0.32 \\
  (within $|l|<3^{\circ}$ and $|l|<3^{\circ}$) &  $V_{\rm lsr}>0$  & 19 & 6.914 &  1.571 &       \\
\hline
\end{longtable} }

{\setlength{\tabcolsep}{2pt}\footnotesize
\begin{longtable}{c|cccc|c}
  \caption{Statistics for the supposed $x_1$ and $x_2$ family sets within $|l|<1.5^{\circ}$
  and $|V_{\rm lsr}|<150$ km s$^{-1}$.} \label{tab7}
  \hline\hline
 Set & quantity  &  number  & mean &  stand.dev. & t-probability \\
     & compared  &          &                 &              &          \\  
\endfirsthead
	\hline
 supposed $x_1$ family set &  $|b|$  &  27  &  0.527  & 0.267  & 0.10  \\
 supposed $x_2$ family set &  $|b|$  &  33  &  0.403  & 0.313  &        \\
  &&&&&\\ 
1. foreside $x_1$ family set  &  $K_{H-K}$  &  17  & 5.267 & 0.973  & 0.10 for set 1 and set 2  \\
2. backside $x_2$ family set  &  $K_{H-K}$  &  15  & 6.000 & 1.390  & 0.29 for set 2 and set 3  \\
3. foreside $x_2$ family set  &  $K_{H-K}$  &  18  & 5.559 & 0.842  & 0.22 for set 1 and set 4  \\
4. backside $x_1$ family set  &  $K_{H-K}$  &  10  & 5.772 & 1.015  &   \\
\hline
\end{longtable}
\vspace{1cm}

\setcounter{figure}{8}
\begin{figure*}
\vspace{-1cm}
  \begin{center}
    \FigureFile(170mm,240mm){}
  \end{center}
  \caption{a. Spectra of the SiO $J=1$--0 $v=1$ and 2 lines for the detected sources.
  The first 8 digit of the 2MASS name and the observing date (yymmdd.d format) are shown.  
  }\label{fig:specrta-a}
\end{figure*}
\setcounter{figure}{8}
\begin{figure*}
\vspace{-1cm}
  \begin{center}
    \FigureFile(170mm,240mm){fig9b.eps}
  \end{center}
  \caption{b. ~ (Continued) }\label{fig:specrta-b}
\end{figure*}
\setcounter{figure}{8}
\begin{figure*}
\vspace{-1cm}
  \begin{center}
    \FigureFile(170mm,240mm){fig9c.eps}
  \end{center}
  \caption{c. ~ (Continued)}\label{fig:specrta-c}
\end{figure*}
\setcounter{figure}{8}
\begin{figure*}
\vspace{-1cm}
  \begin{center}
    \FigureFile(170mm,240mm){fig9d.eps}
  \end{center}
  \caption{d. ~ (Continued)}\label{fig:specrta-d}
\end{figure*}
\setcounter{figure}{8}
\begin{figure*}
\vspace{-1cm}
  \begin{center}
    \FigureFile(170mm,240mm){fig9e.eps}
  \end{center}
  \caption{e. ~ (Continued)}\label{fig:specrta-e}
\end{figure*}
\setcounter{figure}{8}
\begin{figure*}
\vspace{-1cm}
  \begin{center}
    \FigureFile(170mm,240mm){fig9f.eps}
  \end{center}
  \caption{f. ~ (Continued)}\label{fig:specrta-f}
\end{figure*}
\setcounter{figure}{8}
\begin{figure*}
\vspace{-1cm}
  \begin{center}
    \FigureFile(170mm,240mm){fig9g.eps}
  \end{center}
  \caption{g. ~ (Continued)}\label{fig:specrta-g}
\end{figure*}
\setcounter{figure}{8}
\begin{figure*}
\vspace{-1cm}
  \begin{center}
    \FigureFile(170mm,240mm){fig9h.eps}
  \end{center}
  \caption{h. ~ (Continued)}\label{fig:specrta-h}
\end{figure*}
\setcounter{figure}{8}
\begin{figure*}
\vspace{-1cm}
  \begin{center}
    \FigureFile(170mm,240mm){fig9i.eps}
  \end{center}
  \caption{i. ~ (Continued)}\label{fig:specrta-i}
\end{figure*}

\setcounter{figure}{8}
\begin{figure*}
\vspace{-1cm}
  \begin{center}
    \FigureFile(70mm,60mm){fig9j.eps}
  \end{center}
  \caption{j. ~ (Continued)}\label{fig:specrta-j}
\end{figure*}


\begin{thebibliography}{}

\bibitem[Alard~(2001)]{ala01} Alard, C. 2001, \aap, 379, L44

\bibitem[Arad \& Johansson~(2001)]{ara05}
Arad, I., \& Johansson, P. H. 2005, MNRAS, 362, 252

\bibitem[Athanassoula~(1992)]{ath92}
Athanassoula, E., 1992, MNRAS, 259, 328	
	
\bibitem[Babusiaux, Gilmore(2005)]{bab05}
Babusiaux, C. \& Gilmore, G., 2005, MNRAS, 358, 1309


\bibitem[Binney et al.~(1991)]{bin91}Binney, J.,  Gerhard, O. E., Stark, 
A. A., Bally, J., \& Uchida, K. I. 1991,  MNRAS, 252, 210

\bibitem[Binney, Gerhard(1996)]{bin96}Binney, J., \& Gerhard, O. E. 1996, MNRAS, 279, 1005

\bibitem[Binney, Merrifield(1998)]{bin98}Binney, J., \& Merrifield, M. 1998,
Galactic Astronomy (Princeton Univ. Press, Princeton), 588

\bibitem[Bissantz et al.~(2003)]{bis03}	Bissantz, N., Englmaier, P. \& Gerhard, O. 2003, MNRAS, 340, 949

\bibitem[Bissantz et al.~(2004)]{bis04}
Bissantz, N., Debattista, V. P., \& Gerhard, O. 2004, ApJ, 601, L155

\bibitem[Bureau, Athanassoula(1999)]{bur99}
Bureau, M. \& Athanassoula, E. 1999, ApJ, 522, 686B	

\bibitem[Blitz, Spergel(1991)]{bli91}Blitz, L., \& Spergel, D. N.
 1991, ApJ, 379, 631
 
\bibitem[Blitz et al.~(1993)]{bli93} 	
Blitz, L., Binney, J., Lo, K. Y., Bally, J., \& Ho, P. T. P.
1993, Nature 361, 417
 
 \bibitem[Bureau, Athanassoula(2005)]{bur05}	
Bureau, M., \& Athanassoula, E. 2005, ApJ, 626, 159B

 
\bibitem[Caswell et al.~(1981)]{cas81}
Caswell, J. L., Haynes, R. F., Goss, W. M., \& Mebold, U. 
1981, Aust. J. Phys., 34, 333

\bibitem[Carey et al.~(1998)]{car98}
Carey, S. J., Clark, F. O., Egan, M. P., Price, S. D., Shipman, R. F., \& Kuchar, T. A. 1998, ApJ, 508, 721
 
\bibitem[Contopoulos~(1988) ]{con88}
Contopoulos, G. 1988, A\&A, 201, 44

\bibitem[Contopoulos, Grosb\o l~(1989) ]{con89a}
Contopoulos, G., \& Grosb\o l, P. 1989, A\&AR, 1, 261


\bibitem[Contopoulos et al. (1989) ]{con89}
Contopoulos, G. Gottesman, S. T., Hunter, J. H., Jr., \& England, M. N. 	
1989, ApJ, 343, 608
 
\bibitem[Dame et al.~(2001) ]{dam01}	Dame, T. M., Hartmann, D., \& Thaddeus, P. 2001,  \apj, 547, 792

\bibitem[Deguchi et al.~(2000)]{deg00a} 
Deguchii, S., Fujii, T.,  Izumiura, H., Kameya, O., 
Nakada, Y., Nakashima, J., Otsubo, T.,  \& Ukita, N.  
2000, ApJS, 128,  571


\bibitem[Deguchi et al.~(2002)]{deg02} 
Deguchi, S., Fujii, T.,  Miyoshi, M., \& Nakashima, J. 2002, 
PASJ, 54, 61

\bibitem[Deguchi et al.~(2004a)]{deg04a} 
Deguchi, S., Imai, H., Fujii, T., Glass, I., Ita, Y., Izumiura, H., Kameya, O., Miyazaki, A., Nakada, Y., \& Nakashima, J.
 2004, PASJ, 56, 261


\bibitem[Deguchi et al.~(2004b)]{deg04b} 
Deguchi, S., Fujii, T., Glass, I., Imai, H., Ita, Y., Izumiura, H., Kameya, O., Miyazaki, A., Nakada, Y., \& Nakashima, J.
 2004, PASJ, 56, 765


\bibitem[Dwek et al.~(1995)]{dwe95}Dwek, E., 
et al. 1995, ApJ, 445, 716


\bibitem[Englmaier \& Gerhard~(1999)]{eng99}
Englmaier, P. \& Gerhard, O. 1999, MNRAS, 304, 512

\bibitem[Glass et al.~(2001)]{gla01}
Glass, I. S., Matsumoto, S., Carter, B. S.,  \& Sekiguchi, K. 2001, MNRAS 321, 77 

\bibitem[Habing (1987)]{hab87} Habing 1987, in the Galaxy, ed. G. Gilmore \& P. Carswell
(Dordrechr: Reidel), p173

\bibitem[H\"afner et al.~(2000)]{haf00} 
H\"afner, R., Evans, N. W., Dehnen, W., \& Binney, J. 2000, MNRAS, 314, 433

\bibitem[Imai et al.~(2002)]{ima02}
Imai, H., 
et al. 2002, PASJ,  54, L19

   
\bibitem[Izumiura et al.~(1995)]{izu95}Izumiura, H., Deguchi, S., 
Hashimoto, O., Nakada, Y., Onaka, T., 
Ono, T., Ukita, N.,  \& Yamamura, I. 
1995, ApJ,   453,  837 

\bibitem[Jewell et al.~(1991)]{jew91}Jewell, P. R., Snyder, L. E., 
     Walmsley, C. M., Wilson, T. L., 
   \& Gensheimer, P. D. 1991,  A\&A 242, 211

\bibitem[Kaufmann \& Patsis~(2005)]{kau05}
Kaufmann, D. E., \& Patsis, P. A. 2005, ApJ, 624, 693

\bibitem[Kuijken, Merrifield (1995)]{kui95}  
Kuijken, K. \& Merrifield, M.R. 1995, ApJ, 443, L13

\bibitem[Launhardt et al.~(2002)]{lau02}
Launhardt, R., Zylka, R., \& Mezger, P. G. 2002, A\&A, 384, 112
   
\bibitem[Lindqvist et al.~(1992)]{lin92a}Lindqvist, M., 
  Habing, H. J., \& Winnberg, A. 1992, A\&A, 259, 118
  

\bibitem[Messineo et al.~(2002)]{mes02}
Messineo, M., Habing, H. J., Sjouwerman, L. O., Omont, A., \& Menten, K. M. 2002,
A\&A, 393, 115

\bibitem[Miyazaki et al.~(2001)]{miy01}
Miyazaki, A., Deguchi, S., Tsuboi, M., Kasuga, T., \& Takano, S., 2001,
PASJ, 53, 501

\bibitem[Mouhcine, Lan\c con(2002)]{mou02}
Mouhcine, M., \&  Lan\c con, A. 2002, A\&A, 393, 149


\bibitem[Nakada et al.~(1991)]{nak91}Nakada, Y., Deguchi, S., 
Hashimoto, O., Izumiura, H., Onaka, T., 
Sekiguchi, K., \& Yamamura, I. 
1991, Nature, 353, 140

\bibitem[Nakashima,  Deguchi(2003)]{nak03} Nakashima, J., \& Deguchi, S. 2003, PASJ, 55, 203

\bibitem[Nikolaev, Weinberg~(1997)]{nik97}
Nikolaev, S., \&  Weinberg, M. D. 1997, ApJ, 487, 885

\bibitem[Nishiyama et al.~(2005)]{nis05}
Nishiyama, S., Nagata, T., Baba, D., Haba, Y., Kadowaki, R., et al.  
2005, ApJ, 621, L105

\bibitem[Oka et al.~(1998)]{oka98} Oka, T., Hasegawa, T., Sato, F., Tsuboi, M., \& Miyazaki, A. 1998,
ApJS, 118, 455 
    

\bibitem[Paczynski et al.~(1994)]{pat94}
Paczynski, B., Stanek, K. Z., Udalski, A., Szymanski, M., Kaluzny, J., Kubiak, M., Mateo, M., \& Krzeminski, W.
1994, ApJ, 435, L113	


\bibitem[Pichardo et al.~(2004)]{pic04}
Pichardo, B., Martos, M., \& Moreno, E. 2004, ApJ, 609, 144

\bibitem[Price et al.~(2001)]{pri01}Price, S. D., Egan, M. P.,
Carey, S. J., Mizuno, D. R., \& Kuchar, T. A. 2001, AJ, 121, 2819

\bibitem[Regan \& Teuben~(2004)]{reg04}
Regan, M. W., \& Teuben, P. J. 2004, ApJ, 600, 595
   

\bibitem[Sawada et al.~(2004)]{saw04}
Sawada, T., Hasegawa, T., Handa, T., \& Cohen, R. J. 2004, MNRAS, 349, 1167

\bibitem[Schwarzschild~(1979)]{sch79}
Schwarzschild, M. 1979, ApJ, 232, 236

\bibitem[Sevenster et al.~(1997)]{sev97}
Sevenster, M. N., Chapman, J. M., Habing, H., Killeen, N. E. B., \& 
Lindqvist, M. 1997, A\&AS, 122, 79

\bibitem[Sellwood \& Wilkinson~(1993)]{sel93}
Sellwood, J. A. \& Wilkinson, A. 1993, Rep. Prog. Phys. 56, 173

\bibitem[Shiki et al.~(1997)]{shi97}Shiki, S., Ohishi, M., 
\& Deguchi, S. 1997, ApJ,  478, 206


\bibitem[Sjouwerman et al.~(1998)]{sjo98}Sjouwerman, L. O., 
van Langevelde, H. J., Winnberg, A., \& Habing, H. J. 1998, 
A\&AS, 128, 35

\bibitem[Sjouwerman et al.~(2002)]{sjo02}Sjouwerman, L. O., Lindqvist, M. , van Langevelde, H. J.,
\&  Diamond, P. J. 2002, A\&A, 391, 967

\bibitem[Skokos et al.~(2002)]{sko02}
Skokos, Ch., Patsis, P. A., \& Athanassoula, E. 2002, MNRAS, 333, 861

\bibitem[Stark et al.~(2004)]{sta04}
Stark, A. A., Martin, C. L., Walsh, W. M., Xiao, K., Lane, A. P., \& Walker, C. K. 2004, ApJ, 614, L41

\bibitem[Taylor et al.~(1993)]{tay93}
Taylor, G. B.,  Morris, M., \& Schulman, E. 1993, AJ, 106, 1978
 
\bibitem[Tsuboi et al.~(1999)]{tsu99}
 Tsuboi, M., Handa, T., \& Ukita, N. 1999, ApJS, 120, 1

\bibitem[van Langevelde et al.~(1992)]{van92}
van Langevelde, H. J., Brown, A. G. A., Lindqvist, M., 
Habing, H. J., \& de Zeeuw, P. T. 1992, A\&A, 261, L17


\bibitem[Weiner, Sellwood~(1999)]{wei99} 
Weiner, B. J., \& Sellwood, J. A. 1999, ApJ, 524, 112

\bibitem[Whitelock et al.~(1991)]{whi91}
Whitelock, P., Feast, M.,  \& Catchpole, R. 1991, MNRAS 248, 276


\bibitem[Zhao~(1996)]{zha96}
Zhao, H. S. 1996, MNRAS, 283, 149

\bibitem[Zombeck~(1990)]{zom90}
Zombeck, M. V. 1990, Handbook of Space Astronomy and Astrophysics 
(Cambridge University Press, New York) p70 


\end{thebibliography}
\end{document}